# Lasing in Self-Assembled Biophotonic Network

Chaoyang Gong,[1] Zhen Qiao,[1] Song Zhu,[1] Wenjie Wang,[2] Yu-Cheng Chen[1,3,*]

[1] *School of Electrical and Electronic Engineering, Nanyang Technological University, 50 Nanyang Avenue, 639798, Singapore*
[2] *Key Lab of Advanced Transducers and Intelligent Control System of Ministry of Education, Taiyuan University of Technology, 79 Yingze Street, Taiyuan 030024, PR China*
[3] *School of Chemical and Biomedical Engineering, Nanyang Technological University, 62 Nanyang Drive, 637459, Singapore*
*\* yucchen@ntu.edu.sg*

**Abstract:** Mode competition is conventionally regarded as an unwanted phenomenon that induces instability of laser emission. Here, we experimentally demonstrate that the random dynamics behavior of mode competition provides hidden information of the intracavity topology. We introduced the concept of self-assembled biophotonics networks, where human insulin protein with network topology was sandwiched in a Fabry-Perot (FP) cavity. The optical coupling between insulin fibrils and enables strong mode competition. Thanks to the optical feedback provided by the FP cavity, the interaction between the light and insulin network was enhanced. Subtle changes in the network will cause significant changes in laser mode dynamics. By recording the transverse mode dynamics, we reconstruct a graph that reveals the optical correlation in the insulin network. Our results indicate that the graph represents an intrinsic feature and can sensitivity respond to the stimulus. The unclonable and sensitively responsive property of the biophotonic lasing network can potentially be employed in applications in sensing and information encryption.

## 1. Introduction

Networks are ubiquitous in many disciplines of science, ranging from biological signaling[1, 2], protein interactions[3, 4], molecular energy transport[5], social interaction[6], quantum information[7, 8] to optical communication[9]. The complexity of a network can be interpreted by graphs, whose complexity is dictated by the number of nodes and linkage patterns between them[10]. Correspondingly, photonic networks describe a disordered and complex system with a focus on light interactions between distinct elements and connections[11-22]. Light-matter interactions in nano- to-microscale structures can therefore be interpreted into a graph of photonic networks. A branch of studies in photonic networks have illustrated various lasing behavior in different materials and network topology[11-19]. Most of these network structures take advantage of light amplification in a gain material with feedback from multiple scattering centers. Additionally, photonic networks have been extensively applied in optical routing[23-25], waveguiding[26, 27], and light localization[28, 29].

Unlike conventional photonic networks, "biophotonic networks" represent a unique fashion for investigating the light-matter interacting with biological networks, providing a wealth of interesting physical phenomena[13, 19, 28, 30, 31]. Meanwhile, optical microcavities provide opportunities to enhance the light-matter interaction by confining photons for a long time in small volumes, which has been widely employed in nonlinear optics, lasing, and sensing applications[32-40]. Integrating the biophotonic network with a microcavity can potentially detect the tiny heterogeneity in biological structures. As shown in Fig. 1a, when a biophotonic network is confined in a microcavity formed by two highly reflective mirrors, the photons emitting from networks couple with the external Fabry-Perot (FP) cavities, creating a series of "mirror images" of the biological network with enhanced complexity and features.

Here, we investigate the lasing behavior of a self-assembled protein fibril network confined in an FP microcavity, where a topology-dependent transverse mode behavior was discovered. Self-assembled proteins were employed as the biological network due to their significant role in the biological and pathological process. We report the experimental observation of network lasing exhibiting dynamic response and evidence of light coupling in a biophotonic network. Figure 1(a) illustrates the concept of a biophotonic lasing network formed by insulin conjugated with fluorophores, in which insulin can self-assemble into fibrils, networks, and clusters (Fig. 1(b)). As a result of a higher refractive index and higher localized dye concentration in fibrils, transverse laser modes were confined to the fibril network, enabling strong intracavity light-matter interaction. Lasing emissions from self-assembled fibril networks were systematically analyzed according to different network complexity and polarization. Through the process of mode competition, the correlation of the dynamic signal can therefore be resolved by a topological graph at different spectral wavelengths. Subtle changes in the biophotonic network will be amplified and cause a significant impact on the optical coupling efficiencies. The temporal changes of the transverse laser mode were recorded (Fig. 1(c)), through which the inner optical connections in the lasing network are revealed (Fig. 1(d)). Our findings demonstrated that temporal dynamic changes of frequency, wavelength, and intensity at respective spatial localizations represent significant features of a biological network. Finally, we demonstrate the stimulated responses in large-scale biophotonic lasing networks through different lasing wavelengths. Our findings not only provide a novel concept to study dynamic biological networks through amplified interactions on a graph but also pave a new avenue towards designing biological-controlled photonic devices.

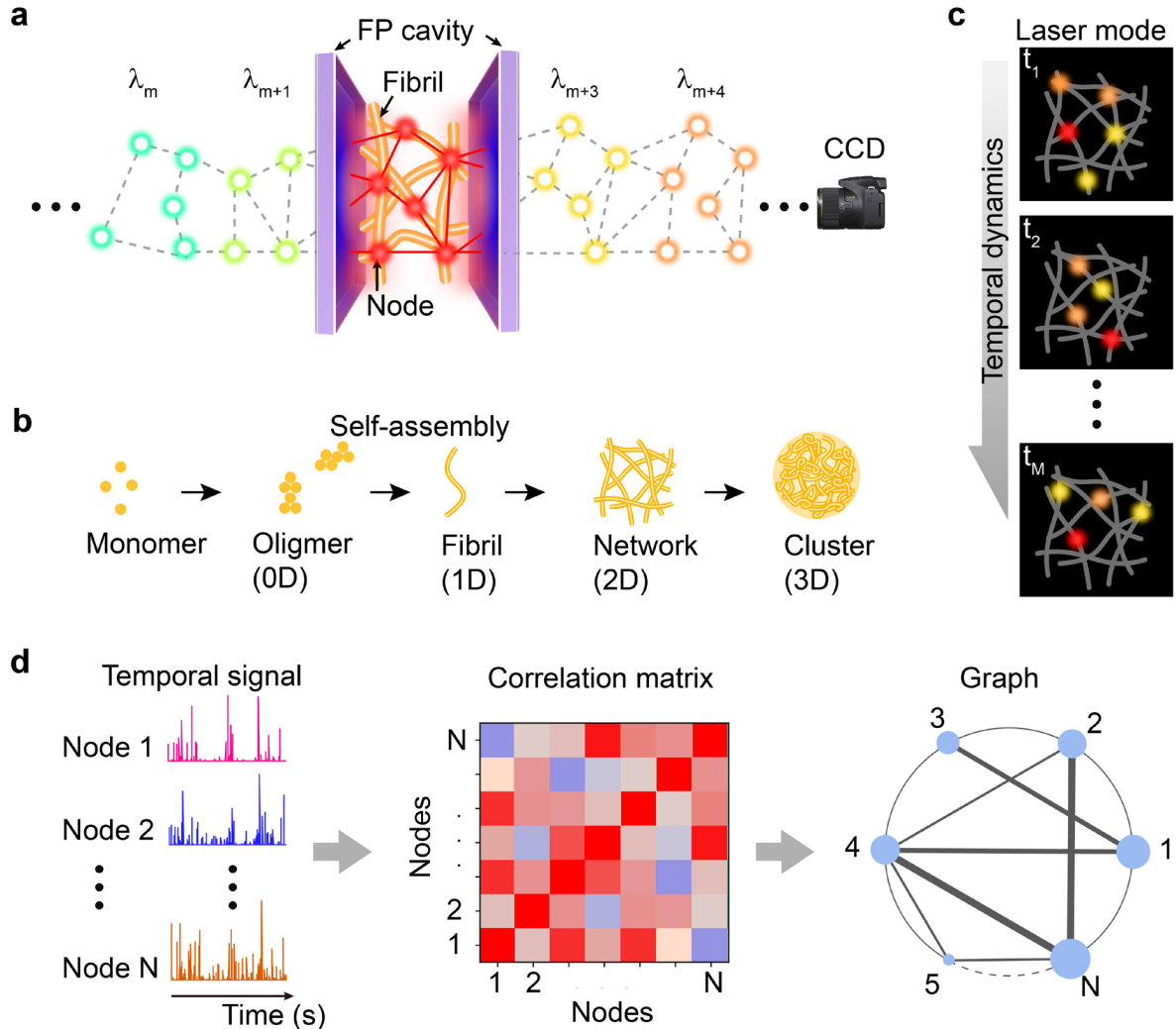

Fig. 1 Concept of the biophotonic lasing network. (a) Schematic illustration of the lasing network. (b) Illustration of insulin monomer self-assembling into different structures. (c) Dynamic laser mode patterns collected at different time. The dots with different colors denote the spatial distribution of laser mode at different wavelengths. (d) Converting the temporal signal of the dynamic laser mode patterns into a graph.

## 2. Results

*2.1 Laser emission from self-assembly structures*

As shown in Fig. 2(a), insulin monomers aggregate into different assembly structures through reaction time [41] (See Methods). For a short incubation time below 2 hours, the monomer insulin molecules begin to form oligomers. Because of the low degree of aggregation, the solution remains clear and no microscopic structures were observed. After 4 hours of incubation, small fibril fragments with a diameter around 1.2 μm can be observed in the turbid solution (Fig. S2). As fibrils continue to aggregate, tiny fibrils assemble into 2-dimensional (2D) network structures after 6 hours of incubation. With a longer incubation time, the network topology structures become more complex and eventually form 3-dimensional (3D) clusters.

Different assembly structures were sandwiched in an FP optical cavity for lasing measurement (See Methods for details). As illustrated in the bottom panel of Fig. 2(a), the distinctive pattern originated from the transverse mode of the FP cavity, which is highly dependent on the assembly structures of insulin. In the absence of fibril, only a single laser mode pattern was observed, corresponding to the fundamental Gaussian mode. Once a single fibril is formed, periodic bright lasing dots (lasing modes) along the fibril could be distinguished. Notice that the excitation beam is only pumped at one end on the insulin fibril; however, a bright lasing spot also appeared on the other end of the fibril, indicating an optical waveguide effect in the fibril. The waveguide effect resulted from the relatively low absorption around 600 nm[42] and the localization of gain molecules on fibrils[13] (Fig. S3). Although the intensity distribution is similar to that observed in a nanowire laser where optical feedback is provided by two end-facets[43, 44], our theoretical calculation confirms that the observed laser emission is the result of FP cavity resonance. The theoretical cavity length is calculated to be 11.8 μm by using $d = \lambda^2 / (2n_{eff} \cdot FSR)$, which roughly agrees with the cavity length (9 μm). Here, $\lambda$ = 600 nm is the lasing wavelength. $FSR$ = 11.4 is the free spectral range (FSR) of emission spectra which can be extracted from Fig. 2(b). $n_{eff}$ = 1.33 is the effective refractive index of the laser mode in the FP cavity (the refractive index of water is used as an approximation). When more fibrils assemble together, a 2D intracavity waveguide network was formed, as shown in Fig. 2(a). Similar to the result in a single fibril, bright spots can be observed in regions outside the pump laser. The light in the intracavity network can couple with each fibril, resulting in a more complex transverse mode pattern. Eventually, fibrils form a compact 3D cluster, where one can observe that the laser mode pattern is the superposition of Ince–Gaussian modes with different orders (Fig. S4).

Lasing spectra from different assembly structures were recorded and compared in Fig. 2(b), corresponding to the structures in Fig. 2(a). Due to strong light-matter interactions, the emission spectra in Fig. 2(b) presented a strong dependence on the intracavity topology structure. Before fibril was formed, the spectrum showed three individual peaks with an FSR of 11.4 nm, corresponding to the longitudinal mode of the FP cavity. Once fibril is formed, the fibril slightly deteriorated the Q-factor of the cavity by inducing absorption and scattering. Consequently, the lasing spectrum only showed two peaks around 600 nm and 610 nm. As the insulin self-assembled into a complex structure (network and cluster), more transverse modes can be excited, resulting in the splitting of an individual longitudinal mode into several peaks. To confirm our findings, a statistical approach was performed by recording the laser spectra from 20 samples (Fig. S5), which confirms the evolution trend of lasing spectra. We further measured the output intensities by increasing the pump energy densities to demonstrate lasing action in Figs. 2(c) and 2(d). The lasing thresholds from four different structures are compared in Fig. 2(c). Due to the high Q-factor of the FP cavity and the large optical gain of the fluorophore, a low lasing threshold of 3.0 μJ/mm$^2$ was achieved before fibril formation. The lasing threshold is continuous to increase with a more complex assembly structure, which indicates a higher intracavity loss with a more complex structure. The highest lasing threshold obtained in our experiment is 9.1 μJ/mm$^2$, which is comparable with the previous biological lasers[45, 46].

In addition, the topology of the assembled structure showed a strong impact on its laser emission polarization. By rotating the polarizer in front of the CCD camera, the lasing signal at different polarization directions could be selectively measured (See methods for details). Here we investigated the lasing polarization of different insulin structures, including monomer and oligomer, fibril network, and cluster. As illustrated in Fig. S6, when no fibril was formed, the polarization of laser emission from the oligomer remains consistent with the pump laser. This phenomenon can be explained by the random distribution of dipole momentum of the rhodamine 6G (R6G) molecules in solution. Dye molecules are excited when the dipole momentum is aligned with the direction of pump laser polarization. In contrast, the polarization of the laser emission from a single fibril is highly dependent on the fibril orientation ($\theta$). As illustrated in Figs. 2(e)-2(f), a maximal lasing intensity was observed when the polarizer is perpendicular to the fibril axis. This result can attribute to a lower intracavity loss of the laser mode with a polarization direction perpendicular to the fibril axis. The polarization behavior of the laser emission from a fibril network reveals the optical coupling behavior, which will be discussed in the next section. As illustrated in Fig. S7, the laser emission from a 3D cluster also showed a linear polarization. However, the polarization direction in clusters differs from each other. We suspect that the polarization direction may be correlated with the inner structure.

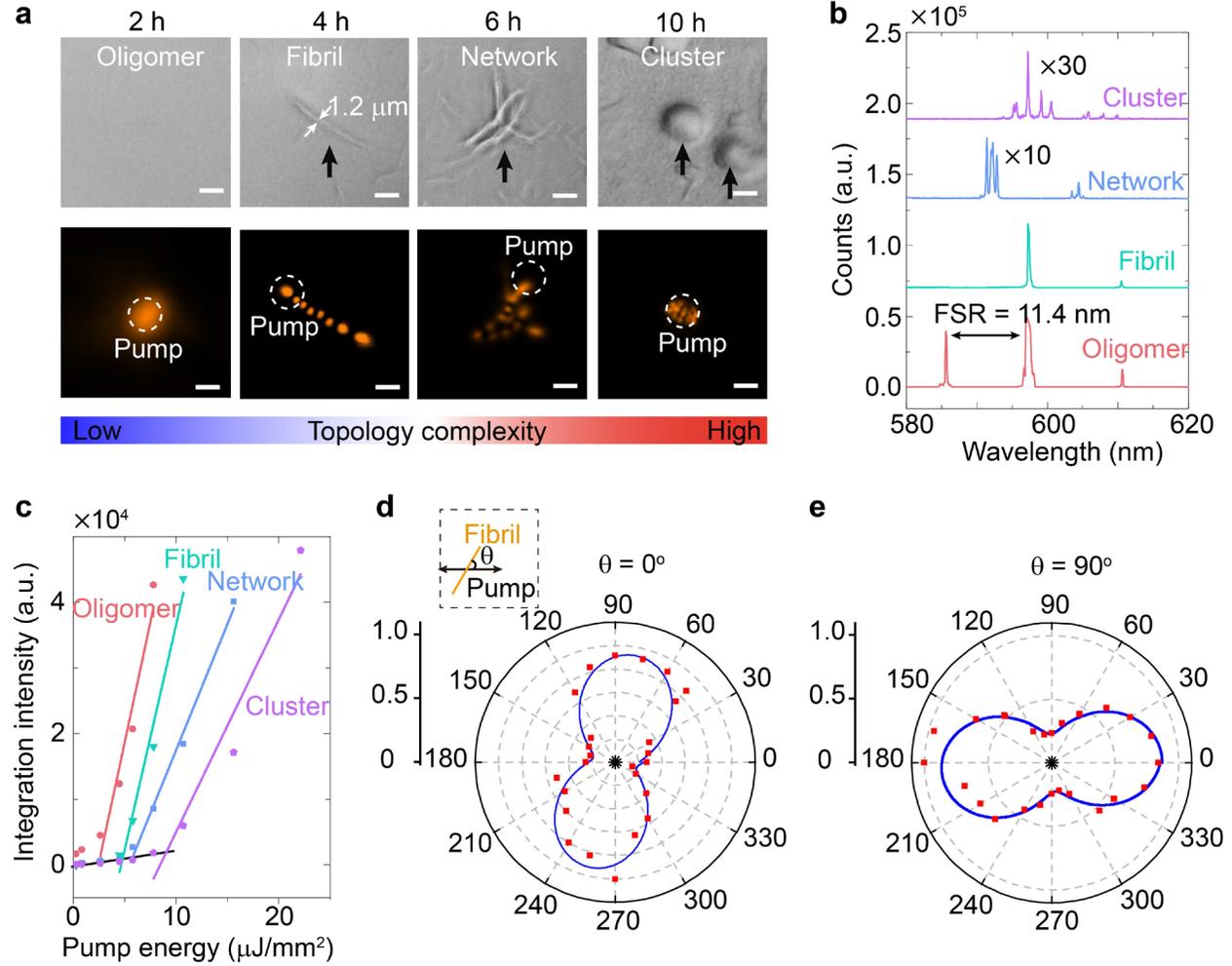

Fig. 2 Laser emission from the self-assembly structure of insulin. (a) Microscope images of insulin self-assembly into different structures (top) and the corresponding laser mode pattern (bottom). The dashed circles denote the location and size of the pump laser. Scale bar: 10 μm. (b) Comparison of the lasing spectra of different assembly structures. c, Lasing intensity from oligomer as a function of pump energy density. The obtained lasing threshold is 3.0 μJ/mm² for oligomer, 4.8 μJ/mm² for fibril, 6.0 μJ/mm² for network, and 9.1 μJ/mm² for cluster. (d)-(e) Polar plot of the laser intensity as a function of the angle between polarization direction of the pump laser and the polarizer, with θ = 0° (d) and 90° (e). The results are fitted with a cosine function (blue curves). θ is defined as the angle between the fibril and the x-axis (inset of d).

## 2.2 Optical coupling in a lasing network

We consider the monochromatic wave of frequency $\omega$ and wave number $k = \omega/c$ confined in the intracavity fibril network, where $\omega$ and $c$ represent the angular frequency and speed of light. As illustrated in Fig. 3(a), the wavenumber $k$ is related to the components through $k^2 = k_x^2 + k_y^2 + k_z^2$, with $k_x$, $k_y$, $k_z$ denoting the wavenumber components along its coordinate axis. In an FP cavity, the component $k_z$ should fulfill the resonance condition $\int_0^L k_z d_z = m\pi$ and the components in the xy-plane $k_{xy}^2 = k_x^2 + k_y^2$ will couple into the fibril network. Consequently, the transverse mode can be regarded as the standing wave pattern of $k_{xy}$ in the fibril network. According to our simulation in Fig. 3(b), the fibril network supports multiple transverse modes and each of them shows individual energy distribution on the fibril network. Note that the energy is distributed randomly in the entire fibril network, rather than localized in a single fibril. The energy distribution between fibrils can be explained by the coupling of $k_{xy}$ in the fibril network.

In order to experimentally confirm that the light couples between fibrils, the polarization of the laser emission from a simple network with two fibrils was investigated (Fig. 3(c)). As illustrated in Figs. 3(d) to 3(e), the polarization directions of the two fibrils show an angular rotation of about 30º on the perpendicular direction of the fibril, which is inconsistent with the result in Figs. 2(e) to 2(f). This result indicates the optical coupling between two fibrils. When reaching the lasing threshold simultaneously, the two fibrils generate laser emission with a polarization direction perpendicular to the fibril axis. As illustrated in Figs. 3(f) and 3(g), the electric field of the laser emission from fibril 1 and fibril 2 is denoted as $E_1$ and $E_2$. A fraction of $E_1$ will be coupled to fibril 2 (denoted as $E_1'$) and similarly, a fraction of $E_2$ will be coupled to fibril 1 (denoted as $E_2'$). The final polarized light will be the superposition of two components, i.e., $E_1 + E_2'$ (for fibril 1) and $E_1' + E_2$ (for fibril 2), resulting in an angular rotation on $E_1$ and $E_2$.

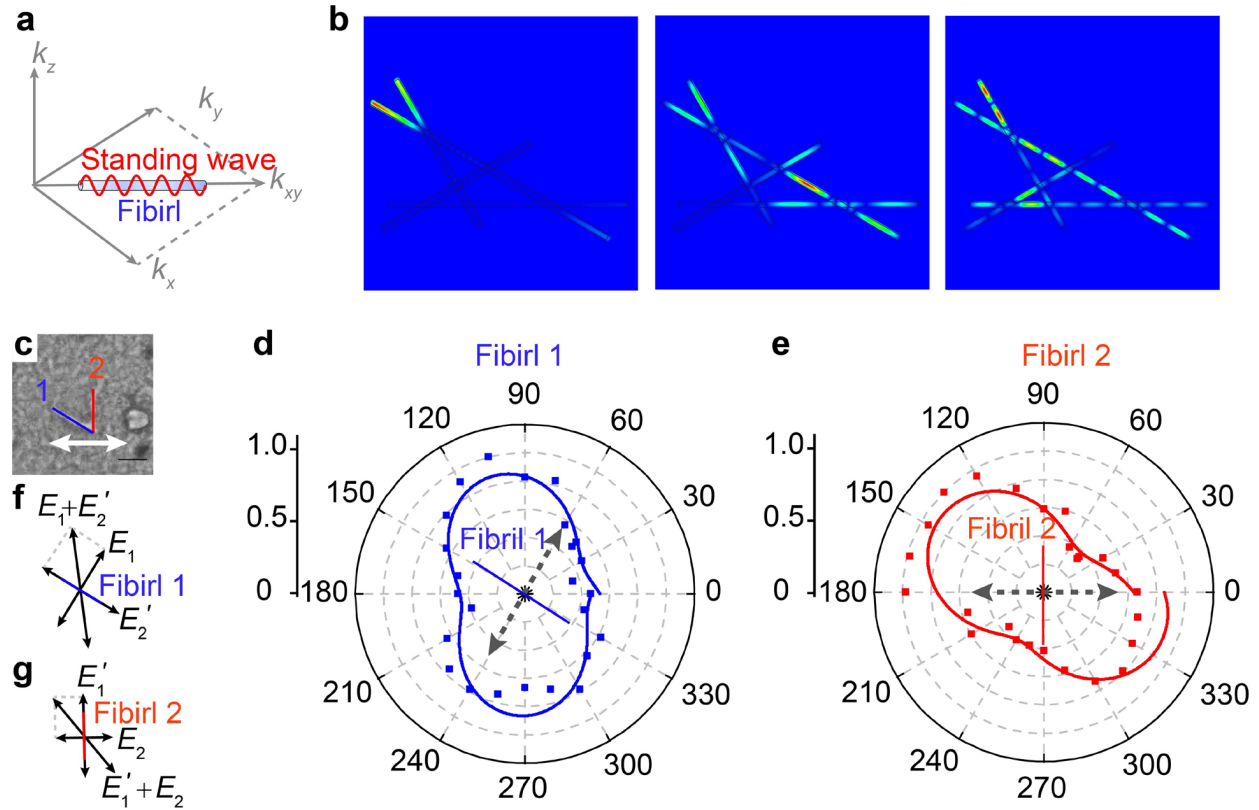

Fig. 3 Optical coupling in a network. (a) Illustration of the optical coupling in a fibril network. (b) Numerical simulation of the transverse mode in a fibril network. (c) Microscopy image of a network form by two attached fibrils. (d)-(e) Comparison of the polarization of the fibril 1 (d) and fibril 2 (e). The orientations of fibril 1 (blue line) and fibril 2 (red line) are also marked. The grey dashed arrows indicate the direction which is perpendicular to the fibril. (f)-(g) Illustration of the superposition of the polarization caused by the optical coupling between fibril 1 (f) and fibril 2 (g).

*2.3 Visualizing the lasing network with a graph*

In this section, we illustrate how the temporal dynamics of lasing behavior in a fibril network can be visualized through a graph. As a proof-of-concept, a fibril network consisting of multiple fibrils was investigated (Fig. 4(a)). According to our findings in the last section (Fig. 3(b)), the lasing network has the capability to support a series of transverse modes sharing similar spatial distribution. When the pump energy density reaches the lasing threshold, mode competition becomes obvious, and significantly distinctive laser mode patterns could be observed at respective time frames. As illustrated in Fig. 4(b), the bright spots that appear on the fibril network indicate the formation of standing waves on the fibril network. According to laser theory, the observed laser patterns will be dominated by the modes experiencing the highest optical gain. This "natural selection" mechanism will selectively mediate the optical coupling in the network, causing energy fluctuation in the temporal domain. By defining each bright spot as a node, the temporal changes of energy flow in the fibril network form an "optical connection" between each node, which can be eventually converted into a graph (Fig. 4(b), right).

Based on the fibril network in Fig. 4(a), we extracted the temporal intensity evolution of each node. Figure 4(c) clearly shows the laser mode pattern recorded at different times. The intensity fluctuation of each node mainly comes from mode competition. Strong evidence of energy flow in the network caused by mode competition can be found in the enlargement. At $t_1 = 14.2$ s, the intensity of node 6 is about 10 times higher than node 4, while at $t_2 = 16.7$ s, the intensity of node 6 is 2.7 times lower than node 4. Note that the stability of the pump laser beam profile in Fig. S8 confirms that the dynamic patterns come from the mode competition rather than the fluctuation of the pump laser. In contrast, the intensity and spatial distribution of the laser modes in a single fibril remain consistent (Fig. S9). This result indicates that mode competition is significantly enhanced when the fibril network is confined in a microcavity. Subsequently, the correlation matrix was calculated as illustrated in the Method. The heatmap in Fig. 4(e) visualizes the correlation matrix between the seven bright spots, indicating a divergence in the correlation between each node. We further converted the correlation matrix into a graph in Fig. 4(f), where the nodes of the graph correspond to the seven bright spots in Fig. 4(a). The edges with different widths denote the correlation. The size of the nodes denotes the weight, which is defined as the sum of the correlation of edges connected to the node.

Our findings demonstrate a graph can be regarded as the fingerprint of a distinctive biophotonic network, which is highly dependent on its topological structure. The graph shows a high reproducibility over time with a fixed network topology. As illustrated in Fig. S10(a), an insulin network consist of 3 fibrils is employed to demonstrate the reproducibility of a graph. The corresponding laser mode pattern is shown in Fig. S10(b). We tested the dynamic laser mode pattern for 6 times., The same method was employed in each measurement (as in Fig. 4). The resulting graph is illustrated in Fig. S11, which confirms a good reproducibility of the graph in a lasing network. Note that the number of fibrils in the network in Fig. S10 is identical to that in Fig. 4. The significant graph in these two situations also confirms a high dependence on the network topology, which may potentially be used to monitor the intracavity changes in a biological network.

Furthermore, we investigated how the network complexity influences the temporal behavior of the lasing network. The temporal evolution in Fig. 4(c) can be viewed as binary information. For instance, if one node is bright, it will be recognized as "1"; otherwise, it will be recognized as "0". Hence, each laser mode pattern can be transformed into a binary number. We induce information entropy to characterize the complexity of the lasing network (See Methods and Fig. S12 for details). As illustrated in Fig. 4(g), with an increasing number of fibrils, the entropy continues to increase, indicating an increasing complexity in the temporal evolution of laser mode pattern. This phenomenon is caused by an increasing number of transverse modes supported by complex fibril networks.

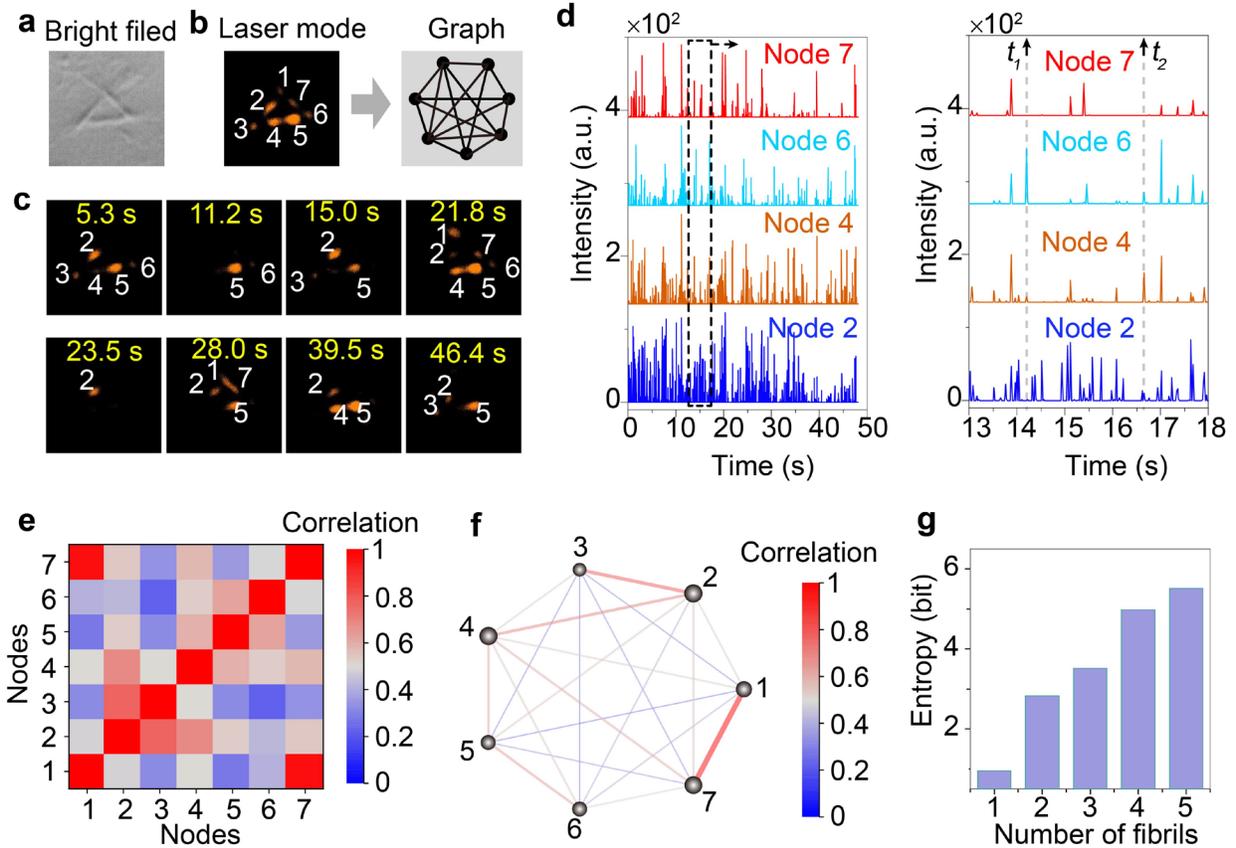

Fig. 4 Visualizing the lasing network with a graph. (a) The optical image of a fibril network. (b) Illustration of converting the laser mode pattern (left) to the corresponding graph (right). (c) Temporal evolution of transverse mode pattern. (d) The intensity of nodes as a function of time (left) and the enlargement (right). (e) The correlation matrix of the lasing network. (f) Visualization of the network in a graph. (g) Entropy as a function as the number of fibrils.

### 2.4 Multiwavelength graph

In addition to temporal changes of the biophotonic lasing network, another important feature is the changes in the spectral domain. Due to the multiwavelength nature of the FP cavity, the lasing network in an FP cavity presents a complex temporal behavior in the lasing spectrum. Herein the temporal behavior was measured by using a spectrograph system. As illustrated in Fig. 5(a), the spectrograph system utilizes a diffraction grating to disperse the laser mode pattern according to its spectral components. The resulting spectral components can be viewed as individual networks with different temporal behavior.

As a proof-of-concept, a biological fibril network with a complex topology was selected and imaged (Fig. 5(b)). Multiple lasing peaks were observed in Fig. 5(c) as a result of a larger Gouy phase shift between each mode in complex topology structures. Figure 5(d) shows the spectral images at four individual lasing peaks (corresponding to the arrows in Fig. 5(c)). The spectral images at different wavelengths show different spatial distribution, implying that transverse modes of the lasing network are highly dependent on the wavelength. This also agrees well with the previous results in Fig. 3(g). Subsequently, according to the definition of the node in Fig. S13, we recorded the temporal evolution of laser patterns at different wavelengths and converted the dynamic patterns into graphs, as presented in Figs. 5(e) to 5(h). Wavelength dependent optical coupling was revealed by the distinctive patterns in the lasing network, providing significant features of the biological fibril network.

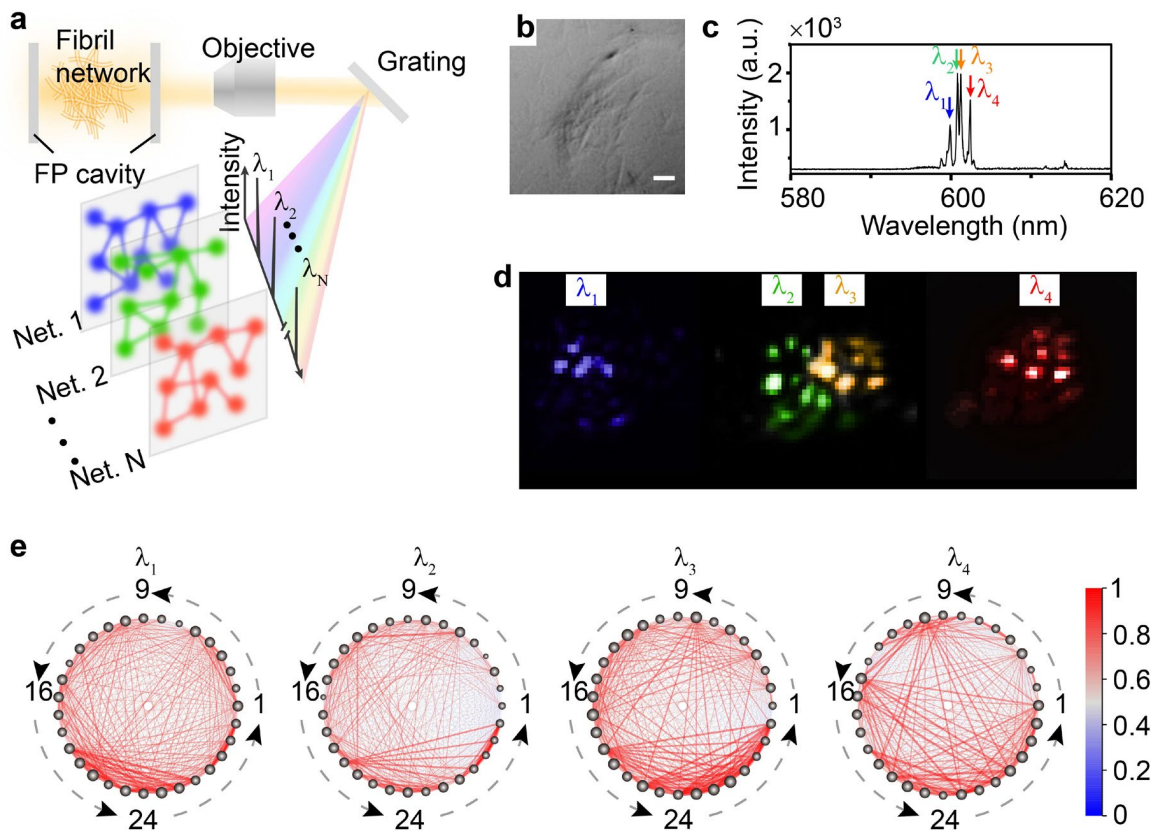

Fig. 5 Hyperspectral imaging of the lasing network. (a) Illustration of the hyperspectral imaging system. (b) Optical image and of a complex network. Scale bar: 10 μm. (c)-(d) the lasing spectrum (c) and the corresponding hyperspectral image at $\lambda_1$ = 600 nm, $\lambda_2$ = 600.8 nm, $\lambda_3$ = 601.2 nm, $\lambda_4$ =602.4 nm (d). (e) Visualization of the networks with multiwavelength graphs at $\lambda_1$, $\lambda_2$, $\lambda_3$, and $\lambda_4$.

*2.5 Stimulus-response of the lasing fibril network*

Finally, we explored the stimulus-response behavior in a biophotonic lasing network. To demonstrate biophotonic lasing network possesses high sensitivity to its surrounding environment, Förster resonance energy transfer (FRET) was employed to induce molecular energy transfer within the fibrils. As illustrated in Fig. 6(a), two organic dyes, i.e., fluorescein isothiocyanate (FITC) and rhodamine B (RhB) are used as FRET pairs. The absorption and emission spectra of the two dyes are illustrated in Fig. 6(b), where the broad spectral overlap ensures an efficient energy transfer. Initially, the fibril networks were labeled with FITC (See Methods for details). The presence of bright fluorescence emission in Fig. S14 indicates the successful binding of FITC molecules on the fibrils. Under pulsed laser excitation, laser emission at 530 nm and 546 nm was observed. A huge redshift of the lasing wavelength with 13.5 nm was observed after adding 0.25 mM RhB in Fig. 6(c). The FITC labeled insulin network showed similar laser behavior as in Fig. 4(c). As a control group, Fig. 6(d) shows the biophotonic laser network images (without RhB) captured at different time frames. In contrast, Fig. 6(e) presents the biophotonic laser network images recorded at different time frames after adding RhB. The transition of the lasing pattern from green to yellow is caused by the shifting of the gain profile. Interestingly, we discovered that the spatial distribution of the laser mode changes significantly after inducing RhB (Fig. S15). The reasons are two folds. Firstly, the wavelength shift induces a phase shift, resulting in energy redistribution among transverse modes. Secondly, the modes with the highest gain emerge during mode competition. As such, the shifting of the gain profile disturbs the original equilibrium in mode competition, selecting a new family of transverse modes with the highest gain profile. The dynamic laser mode patterns before and after inducing RhB were subsequently transferred into graphs. As illustrated in Fig. 6f and Fig. 6g, the graphs clearly show that RhB induced changes in the connection of each node. In other words, changes in the gain profile will be amplified by mode competition, resulting in significant changes on a graph. The findings here demonstrate that such lasing network is very sensitive to subtle changes or stimuli within the biological network.

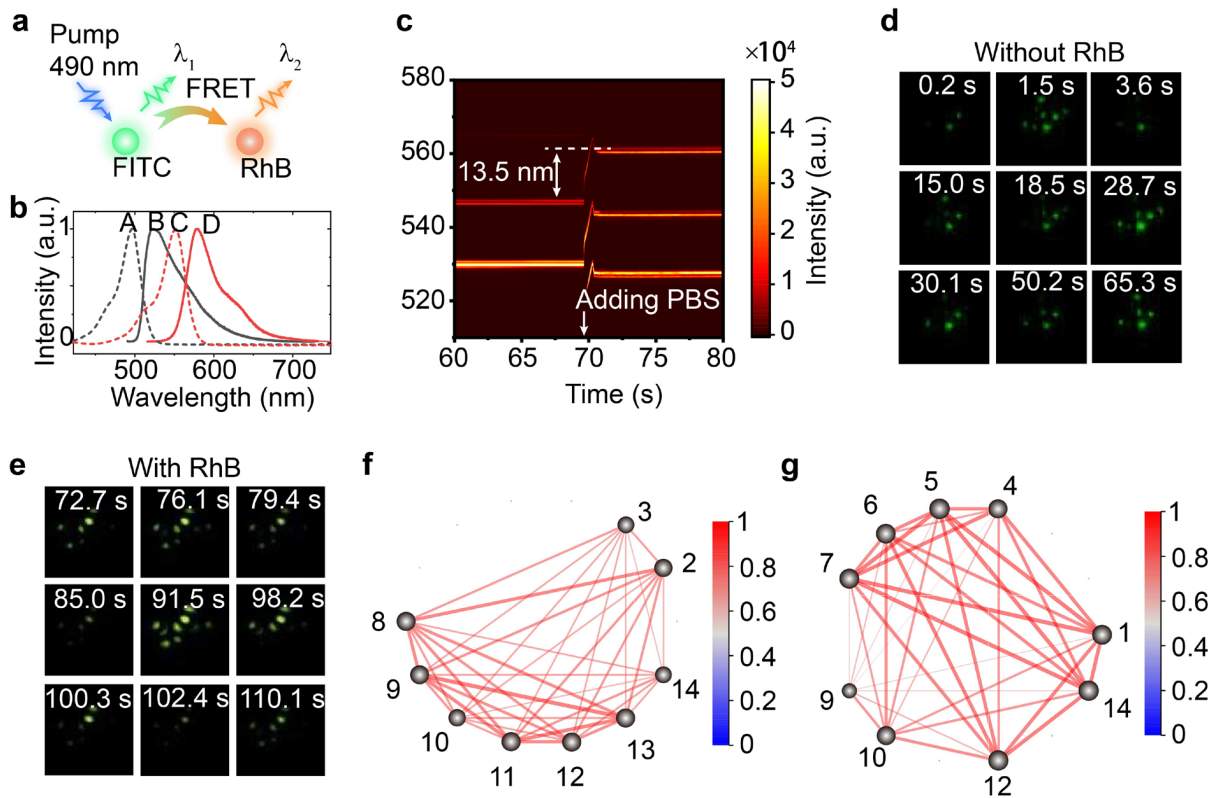

Fig. 6 Controlling the lasing network with Förster resonance energy transfer. (a) Conceptual illustration of the Förster resonance energy transfer (FRET) in the network. (b) Absorption and emission spectra of FITC and RhB. A and B stand for the absorption and emission spectrum of FITC. C and D stand for absorption and emission spectrum of RhB. (c) Temporal recording of the spectra evolution of the lasing network after inducing RhB. (d)-(e) Comparing of the temporal evolution of laser mode pattern before (d) and after (e) inducing RhB. (f)-(g) Comparing of the visualized network graph before (f) and after (g) inducing RhB.

## 3. Discussion and conclusion

Microcavities provide opportunities to confine light within a miniature structure, producing extremely interesting optical phenomena including lasing, four-wave mixing, stimulated Raman scattering, etc. Integrating the biological system into the microcavity yields enhanced light-matter interaction, which is capable to distinguish tiny intracavity changes in molecules, cells, tissues, and living organisms. In this study, we investigate the laser mode dynamics in a fibril network confined in an FP microcavity. By regarding the energy fluctuation as "energy flow", the optical connection in each node can be reconstructed by calculating the correlation of the temporal signal. Our result shows that the laser mode pattern strongly depends on the network topology and the laser mode dynamics is highly sensitive to the stimulation from the medium.

Another important concept which we have introduced in this study is the distinctive graph from a biophotonic network. In nature, biological structures and interactions are entirely unclonable and highly unique. The ability to detect multiple dynamic changes simultaneously and to characterize the huge heterogeneity among complex network interactions is therefore very challenging. Thanks to the strong optical feedback provided by the optical cavity, a small change in the gain induced by the underlying biological processes/interaction is significantly amplified, leading to a drastic change in the lasing output characteristics on a graph. The multiple parameters which network lasing can provide through cavity amplification are overwhelming, including lasing spectrum, peak wavelengths, intensity, threshold, emission directionality, spatial mode distribution, and polarization. The concept of biophotonic lasing thus offers a highly unclonable and distinctive feature for biological networks.

In a nutshell, we would like to discuss a few possible areas that may benefit from the concept of biophotonic lasing network. Firstly, our results suggest that the unique graph can sensitivity respond to the surrounding medium. This interesting property of the graph can be further employed for detecting the tiny structural changes in the biological system like cellular interactions with the extracellular matrix. With the integration of hyperspectral imaging systems, multi-dimensional information of biological system can be revealed at different wavelengths, providing a more accurate and quantitative method for bioassay. Secondly, the unclonable properties of the graph can be further employed in information encryption. The information encoded into the optical connection can be read out by converting the dynamic laser mode pattern into a graph. Thirdly, the graph may also provide deep insights into how biomolecules interact with and modulate the laser light, which lays the foundation for future development of novel bioinspired photonic devices that take advantage of unique self-recognition, self-regulation, of biomolecules/cells to modulate photonic devices.

## 4. Methods

### 4.1 Optical system setup

The experimental setup is illustrated in Fig. S1. Two highly reflective mirrors with a reflectivity of 99.9% (560-730 nm or 500-700 nm) were used to form an FP cavity. The mirrors with a high reflection band in 560-730 nm were used for R6G lasing experiment, while the 500-700 nm mirrors were used for FITC lasing experiment. The insulin with different assembly structures was put on the bottom mirror and was sandwiched in the FP cavity. The two mirrors were separated by glass beads with a diameter of 9 μm. A Nikon Ni-E upright microscope with 20× objective was used for excitation and signal collection. A pulsed laser (EKSPLA PS8001DR, 50Hz, 5 ns pulse width) with an optical parametric oscillator (OPO) was used as an optical pump. The laser emission from OPO was focused on the bottom mirror, forming a spot with a diameter of 14 μm. The laser emission was split by a beam splitter and sent to a spectrometer (Andor Kymera 328i) and a scientific CMOS camera (Andor Zyla SCMOS) for spectrum and image collection. For FRET measurement, the CMOS camera was replaced with a color charge-coupled device (CCD) to monitor the color change of the laser mode pattern. The frame rate of the camera was set as 50 frames per second, which is identical to the repetition rate of the pump laser. Hence, each frame collected from the camera corresponds to the laser emission from one pump pulse. A broadband light-emitting diode (LED) was also monitored on the microscope for fluorescence excitation.

### 4.2 Self-assembly of insulin into different structures

The insulin solution with a concentration of 1mM was prepared by dissolving the insulin powder (Sigma, No. 91077C) in acetic acid solution (20%, pH 2.0) containing 100 mM NaCl. Then, the insulin solution was incubated at 50 ºC with a controlled incubation time (2 to 10 hours) to obtain different assembly structures.

### 4.3 Labeling the insulin with Rhodamine 6G

The gain solution with a concentration of 5 mM was prepared by dissolving the Rhodamine 6G (R6G) powder (Sigma, No. R4127) in phosphate buffer saline (PBS) containing 50% ethanol. The incubated insulin solution was mixed with the gain solution with a volume ratio of 1:1 and incubated at room temperature for 30 min. The final solution is ready for the laser measurement.

### 4.4 Labeling the insulin with FITC

After self-assembly, the insulin solution was centrifuged, washed three times with triethylammonium bicarbonate (TEAB) (Sigma, No. T7408), and are ready for labeling. The FITC stock solution with a concentration of 50 mM was prepared by dissolving the NHS-Fluorescein (Thermo Scientific, No. 46410) powder into dimethyl sulfoxide (DMSO). The FITC working solution with a concentration of 5 mM was freshly prepared by diluting the stock solution with

TEAB. The FITC working solution was mixed with the washed insulin solution with a volume ratio of 1:1 and incubated at room temperature for 30 min. The final solution is ready for the laser measurement.

*4.5 Transfering insulin to FP cavity*

The labeled insulin was transferred to the bottom mirror with a pipette. Because of the higher density than water, the assembled insulin will quickly settle on the bottom mirror. The silica beads with a diameter of 9 μm were induced on the bottom mirror to serve as spacers. Then, the top mirror was put on the bottom mirror, forming an FP cavity with a fixed cavity length. The FP cavity loaded with insulin can be further used for laser measurement.

*4.6 Fluorescence image measurement*

For fluorescence image measurement of R6G labeled fibril, the insulin solution was centrifuged, washed three times with PBS. A bandpass filter with a center wavelength of 560 nm and a bandwidth of 40 nm was employed to select the green pump from the LED.

For fluorescence image measurement of FITC stained fibril, the insulin solution was centrifuged, washed three times with TEAB. a bandpass filter with a center wavelength of 480 nm and a bandwidth of 30 nm was used instead.

*4.7 Calculating the correlation matrix*

As illustrated in Fig. S12, the dynamic transverse mode pattern can be regarded as a temporal signal. Considering two nodes (node *i* and node *j*) with the intensity evolution denoted as a row vector $X_i = [x_{i,1}, x_{i,2}, \cdots x_{i,M}]$ and $X_j = [x_{j,1}, x_{j,2}, \cdots x_{j,M}]$, respectively. *M* is the sample size, i.e., the number of laser mode patterns. Then, the correlation coefficient of the two nodes is defined as

$$r_{i,j} = \frac{\sum_{m=1}^{M}(x_{i,m} - \overline{x_i})(x_{j,m} - \overline{x_j})}{\sqrt{\sum_{m=1}^{M}(x_{i,m} - \overline{x_i})^2 \cdot \sum_{m=1}^{M}(x_{j,m} - \overline{x_j})^2}} \quad (1)$$

For a network with the node number of *N*, the correlation matrix *R* can therefore be written as

$$R = \begin{pmatrix} r_{1,1} & r_{1,2} & \cdots & r_{1,N} \\ r_{2,1} & r_{2,2} & \cdots & r_{2,N} \\ \cdots & \cdots & \cdots & \cdots \\ r_{N,1} & r_{N,2} & \cdots & r_{N,N} \end{pmatrix} \quad (2)$$

*4.8 Converting the correlation matrix into a graph*

The correlation matrix illustrated in Eq. 2 can subsequently be converted into a graph. The edges in the graph denote the correlation coefficient ($r_{i,j}$) between node *i* and *j*. The size of the nodes *i* denotes the weight, which is defined as $w_i = \sum_{j=1}^{N} r_{i,j}$.

*4.9 Calculating the information entropy*

Detailed procedural for calculating the information entropy is illustrated in Fig. S12. Briefly, the temporal signal is firstly obtained from the dynamic laser mode pattern. Then, the temporal signal is converted into a binary signal with a given threshold. Hence, each laser mode pattern corresponds to a binary number. Finally, the information entropy is defined as the expected value of information content, which can be calculated using $H(X) = -\sum_{m=0}^{M} p(T_m) \log_2 p(T_m)$. Here, $p(T_m)$ is the probability of obtaining the binary number $T_m$ from the *m*th pattern.